\documentclass{PoS}
\usepackage{lineno}
\usepackage{array}
\usepackage{booktabs}
\usepackage{multirow}

\usepackage{subcaption}
\usepackage{graphics}
\usepackage{floatrow}

\usepackage{enumitem}

\title{Search for high-energy neutrinos from AGN cores}
\ShortTitle{Search for high-energy neutrinos from AGN cores}
\author{
The IceCube Collaboration$^{\dagger}$\\
{$^{\dagger}$ \itshape \href{http://icecube.wisc.edu/collaboration/authors/icrc19_icecube}{http://icecube.wisc.edu/collaboration/authors/icrc19\_icecube}}\\
E-mail: \email{federica.bradascio@desy.de}
}

\abstract{IceCube is a cubic-kilometer Cherenkov telescope operating at the South Pole. In 2013 IceCube discovered high-energy astrophysical neutrinos and has more recently found compelling evidence for a flaring blazar being a source of high-energy neutrinos. However, as blazars can only be responsible for a small fraction of the observed neutrino flux, the sources responsible for the majority of the detected neutrinos remain unknown. In this work, we explore the possibility that the observed neutrino flux is produced in the cores of Active Galactic Nuclei (AGN). Various models have predicted neutrino emission from the accretion disks of AGN. According to these models, the neutrino luminosity would not depend strongly on either the orientation or other parameters of the relativistic jet. Both jetted and non-jetted AGN could contribute to the neutrino flux. We perform a stacking analysis to test for correlation between various sub-populations of AGN and high-energy neutrinos using a decade of IceCube data. We select AGN based on their radio emission, infrared color properties, and X-ray flux using the NVSS, AllWISE, ROSAT and XMMSL2 catalogs. We use the accretion disk luminosity, estimated from the observed soft X-ray flux, to weight the contribution of selected galaxies to the neutrino signal.\\

\vspace{4mm}
{\bfseries Corresponding authors:}
\speaker{Federica Bradascio}$^{1}$\\
{$^{1}$ \itshape DESY, D-15735 Zeuthen, Germany}\\}

\FullConference{36th International Cosmic Ray Conference -ICRC2019-\\ 
July 24th - August 1st, 2019\\
Madison, WI, U.S.A.}

\begin{document}

\section{Introduction}
\label{sec:intro}
IceCube is a cubic-kilometer Cherenkov telescope operating at the South Pole, detecting neutrinos of all flavors with energies from tens of GeV to several PeV \cite{Aartsen:2016nxy}. In 2013 IceCube discovered high-energy astrophysical neutrinos \cite{1242856} and has more recently found compelling evidence for a blazar being a source of high-energy neutrinos \cite{eaat1378, Aartsen:2018_TXSmultimessenger}. However, the sources responsible for the emission of the majority of the detected neutrinos are still unknown. 
Active Galactic Nuclei (AGN) are considered promising potential sites for high-energy neutrino production as they are the most powerful emitters of radiation in the known Universe. 
They can accelerate protons up to the observed maximum energies of $\sim 10^{20} - 10^{21}$ eV and are surrounded by high-intensity radiation fields where photo-nuclear reactions with subsequent neutrino production can occur.
The most popular scenario for neutrino production is in relativistic jets, as these jets dominate the gamma-ray sky. However, it has been shown that gamma-ray blazars can only be responsible for a small fraction of the observed neutrino flux \cite{Aartsen_2017}. 
Alternatively, sites of hadronic acceleration must avoid over-producing detectable gamma-ray emission, and in this contest obscured sources are favoured \cite{Murase:2016_HiddenSources}. High-energy neutrino emission from the cores of AGN would satisfy this constraint with both the $pp$ \cite{PhysRevD.89.123005} and $p\gamma$ \cite{PhysRevLett.66.2697, Kalashev2015} scenarios.

In this work we explore the possibility that the high-energy neutrinos are produced in the accretion disk that surrounds the super-massive black hole (SMBH) at the heart of an AGN. A maximum likelihood stacking analysis is presented to search for a cumulative signal from multiple sources, selected based on their radio emission, infrared (IR) color properties and X-ray flux. 

\section{Theoretical models}\label{sec:mod}
In this paper,  two models for neutrino production in AGN cores will be tested:
\begin{enumerate}
    \item \textbf{Neutrinos from AGN with Shakura-Sunyaev accretion disk}\\
    AGN generally have a geometrically thin, optically thick disk, as described by the Shakura-Sunyaev model \cite{SS73}. 
    The disk is hot and emits thermal radiation producing a prominent feature in the observed AGN spectra at $\sim 10$~eV, usually referred as the ``big blue bump''.
    Protons accelerated in the vicinity of the SMBH horizon can interact with the "blue  bump'' photons via the $p\gamma \rightarrow n\pi^{+}$ reaction, generating high-energy neutrinos.

    \item \textbf{Neutrinos from LLAGN with RIAF}\\
    Low Luminosity AGN (LLAGN) do not have standard disks, since their spectra show no ``blue bump'' \cite{doi:10.1146/annurev.astro.45.051806.110546}.
    Instead, they are expected to have Radiative Inefficient Acceleration Flows (RIAFs) which are formed when the mass accretion rate into the SMBH is relatively small.
    In the turbulent plasma of the RIAFs, cosmic ray protons may be accelerated via stochastic processes or by magnetic reconnection \cite{Kimura_2015}. These protons then interact with other nucleons ($pp$ interactions) and photons ($p\gamma$ interactions) in the flow, generating neutrinos. 
\end{enumerate}

\noindent In both cases, we can assume that the neutrino luminosity is proportional to the accretion disk luminosity in UV, since at this wavelength the accretion disk emission has its peak. However, a tight relation between UV and X-ray luminosity has been observed in AGN, indicating a connection between the primary radiation from the disk and the X-ray emission from the hot-electrons corona \cite{Lusso_2016, Steffen_2006}. Therefore, the measured soft X-ray flux will serve as an estimate for the expected neutrino flux.

%%%%%%%%%%%%%%%%%%%%%%%%%%%%%%%%%%%%%%%%%%%%
%           Source selection               %
%%%%%%%%%%%%%%%%%%%%%%%%%%%%%%%%%%%%%%%%%%%%
\section{Source selection}\label{sec:cat}
To test the two AGN core models, we want to select a large, clean sample of radio galaxies with X-ray flux info, that is free from blazar contamination. Three samples are created through the cross-correlation of X-ray, radio and infrared catalogues: radio--selected AGN, IR--selected AGN and LLAGN. The correlation among catalogues consists in a positional match performed using the \texttt{extcat}  code\footnote{https://github.com/MatteoGiomi/extcats}. 

The primary X-rays catalogues used for cross--matching are the ROSAT All-sky Survey (2RXS; \cite{refId0}) and the second release of the XMM-Newton Slew Survey (XMMSL2\footnote{https://www.cosmos.esa.int/web/xmm-newton/xmmsl2-ug}). 
However, rather than directly using these two catalogues, we use the ones modified by \cite{10.1093/mnras/stx2651}. They provide 106,573 and 17,665 X-ray sources from the 2RXS and XMMSL2 surveys respectively, which have been matched with AllWISE IR counterparts \cite{Wright_2010}, covering $\sim 95\%$ of the extragalactic sky ($|b|> 15^{\circ}$).
The radio-- and IR--selected AGN samples are compiled by cross-matching the X-ray catalogues with the NRAO VLA Sky Survey (NVSS; \cite{Condon_1998}) radio catalogue and the AllWISE catalogue respectively.  
In the end, all three catalogues are cross-matched with the 3LAC \emph{Fermi}-LAT catalogue \cite{Ackermann_2015} to remove blazars from the final samples. In this analysis we focus only on radio galaxies, for which we expect the emission to come mainly from the accretion disk. 
Table~\ref{tab:catalogues} shows the three AGN samples created for this work, the original catalogues from where they are derived and the weighting scheme used in the analysis.

\begin{table}[h]
    \caption{\label{tab:catalogues}Properties of the AGN samples created for the analysis. The surveys used for the cross-match to derive each sample, the final number of selected sources and their weight are listed.
    }
    \centering
    \resizebox{\columnwidth}{!}{%
    \setlength{\belowrulesep}{5pt}
    \begin{tabular}{r c c c}
        \toprule
        & \textbf{{Radio--selected AGN}} &  \textbf{{IR--selected AGN}} &  \textbf{LLAGN}\\
        \midrule[\heavyrulewidth]
        \midrule[\heavyrulewidth]
        \multirow{2}{*}{\textbf{Matched catalogues}} & \textsc{NVSS} + & \textsc{AllWISE} + & \multirow{2}{*}{\textsc{AllWISE} + \textsc{2RXS}}\\
        & \textsc{2RXS} + \textsc{XMMSL2} & \textsc{2RXS} + \textsc{XMMSL2} &  \\
        \textbf{Nr. of selected sources} & $13,927$ &   $52,835$ & $25,648$\\
        \textbf{Weight} & X-ray flux &  X-ray flux & X-ray flux $\times$ Seyfertness\\
        \bottomrule
    \end{tabular}
    }
\end{table}

%%%%%%%%%%%%%%%%%%%%%%%%
% Radio-Selected AGN %
%%%%%%%%%%%%%%%%%%%%%%%%
\subsection{Radio--selected AGN}
We correlate the NVSS and 2RXS catalogues, keeping all sources whose radio and X-ray positions differ by less than 60 arcsec. This selection criterion matches 14,760 NVSS sources with 12,851 2RXS sources. $6\%$ of X-ray objects have more than one radio counterpart, and for this case the closer NVSS source to the X-ray counterpart is chosen. 
The X-ray sources with a radio counterpart are then cross-matched with the 3LAC catalogue in order to remove blazars, using the 95\% source position error of the gamma-ray sources as search-radius. The gamma-ray BL Lacs, FSRQs and blazars of uncertain type obtained from the cross-match are thus removed from our sample. 

We repeat the same procedure using the XMMSL2 catalogue: we first cross-match it with the NVSS sources using a search-radius of 60 arcsec and then we remove the 3LAC blazars. Before combining the NVSS/2RXS and NVSS/XMMSL2 samples, we remove duplicated X-ray sources (i.e. XMMSL2 sources already included in 2RXS) and convert the X-ray fluxes to the common 0.5-2 keV energy range. The final radio--selected AGN sample contains 13,927 sources with an estimated contamination of only $\sim 5\%$ and an efficiency of $\sim 94\%$, covering $\sim 56\%$ of the sky. 

%%%%%%%%%%%%%%%%%%%%%
% IR-Selected AGN %
%%%%%%%%%%%%%%%%%%%%%
\subsection{IR--selected AGN}
\label{sec:ir_AGN}
Another way to select AGN is to use the IR color-color diagrams, as different classes of objects appear in different regions according to the shape of their spectral energy distribution. We start from the 2RXS and XMMSL2 catalogues provided in \cite{10.1093/mnras/stx2651} and we use the distribution of the AllWISE counterparts to select only AGN, using the color-color diagrams from \cite{Wright_2010}. 
In addition, since some of the sources in the 2RXS catalogue have counterparts in the VERONCAT \cite{VERONCAT} catalogue, and are thus classified as AGN, BL Lac or quasars, we overlap our catalogue with these classified sources and apply cuts to isolate the AGN sources. After removing the 3LAC blazars, we are left with 52,835 sources. 

%%%%%%%%%
% LLAGN %
%%%%%%%%%
\subsection{LLAGN}
\begin{figure*}[t!]
    \centering
    \begin{subfigure}[t]{0.5\textwidth}
        \centering
        \includegraphics[width=\textwidth,clip]{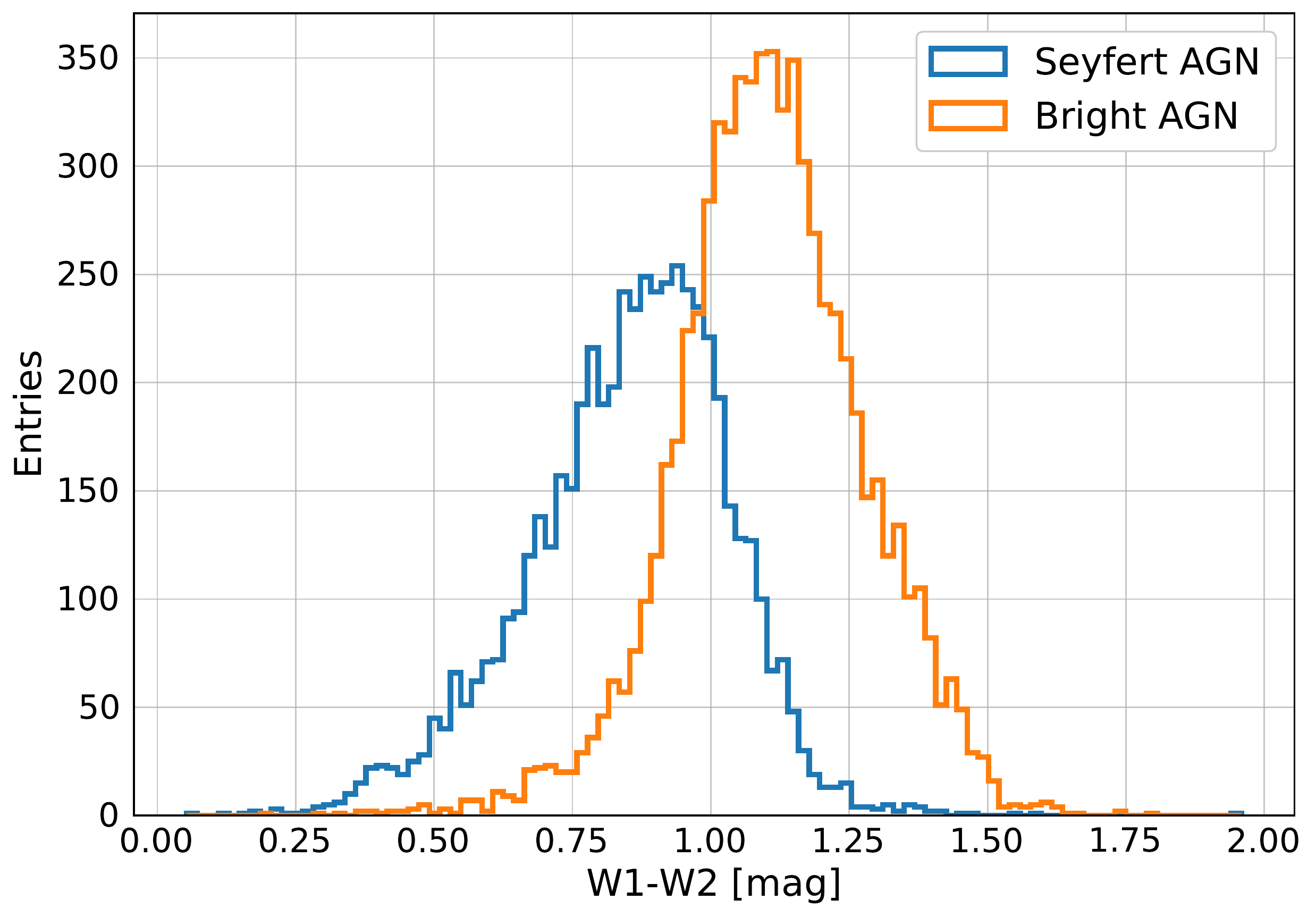}
        \caption{\label{fig:sey_distr}}
    \end{subfigure}%
    ~ 
    \begin{subfigure}[t]{0.5\textwidth}
        \centering
        \includegraphics[width=\textwidth,clip]{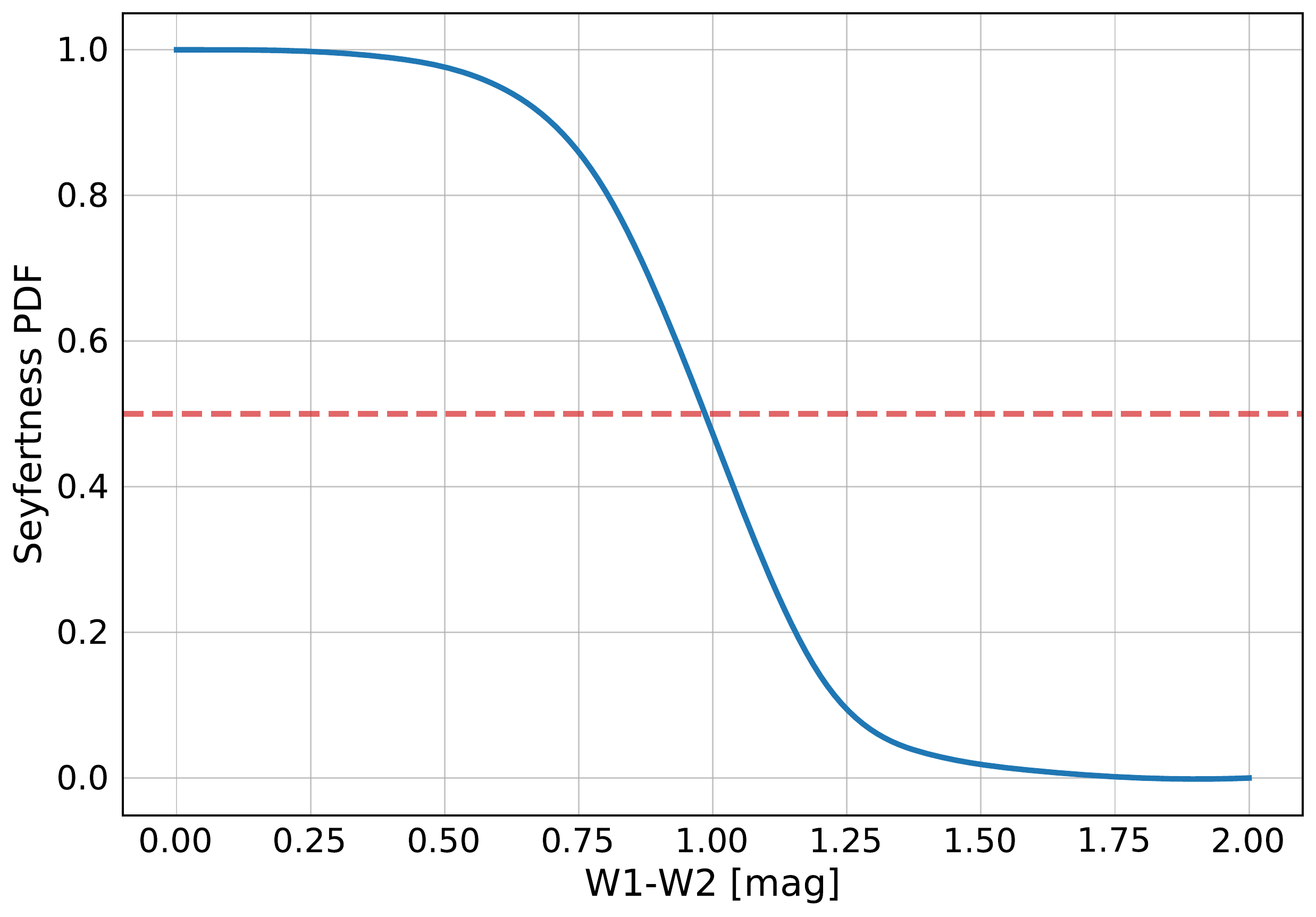}
        \caption{\label{fig:sey_pdf}}
    \end{subfigure}
    \caption{\label{fig:seyfertness}Seyfertness PDF definition. The left panel (\ref{fig:sey_distr}) shows the distributions of the 2RXS sources classified as Seyfert or bright AGN in the VERONCAT as a function of the W1-W2 color-magnitude. The right panel (\ref{fig:sey_pdf}) is the Seyfertness PDF function derived from Figure \ref{fig:sey_distr}; the dashed-red line shows the cut used for the creation of the LLAGN sample.}
\end{figure*}

To build this catalogue, we start from the 2RXS sources with AllWISE counterparts. We remove the 3LAC blazars and apply the same cuts as in section \ref{sec:ir_AGN}. Using the VERONCAT classification, it is possible to separate the bright AGN from the Seyfert galaxies in the W1-W2 space, where W1 and W2 are the 3.4 $\mu$m and 4.6 $\mu$m luminosity respectively (see Figure \ref{fig:sey_distr}). Based on these distributions, a ``Seyfertness'' probability distribution function (PDF) is defined as:
\begin{equation}
\label{eq:sey_pdf}
    \mathrm{Seyfertness} = \frac{P(\mathrm{S})}{P(\mathrm{S})+P(\mathrm{B})},
\end{equation}
where $P(\mathrm{S})$ and $P(\mathrm{B})$ are the probability of being a Seyfert or a bright galaxy respectively. Figure \ref{fig:sey_pdf} shows the Seyfertness PDF as function of W1-W2. Using this function, we assign to each source a Seyfertenss PDF between 0 and 1. In the final sample we only include sources with Seyfertness~$\geq 0.5$, since at this value we have the best trade-off between efficiency (77\%) and contamination (21\%) of the selection. The final LLAGN sample contains 25,648 sources covering $\sim 95\%$ of the extragalactic sky, which will be weighted by their X-ray flux and Seyfertness. 

%%%%%%%%%%%%%%%%%%%%%%%%%%%%%%%%%%%%%%%%%%%%
%               Analysis                   %
%%%%%%%%%%%%%%%%%%%%%%%%%%%%%%%%%%%%%%%%%%%%
\section{Analysis} \label{sec:met}

A stacking analysis is performed using 10 years of IceCube data \cite{Tessa:2019icrc}. The analysis uses an unbinned maximum likelihood ratio test, which gives the significance of an excess of neutrinos above  background expectations for a given direction \cite{PS_7year}. 
Both a signal and a background PDF enter into the likelihood function (equation 2 in \cite{PS_7year}). 
The background PDF is constructed directly from data under the assumption that the dataset is background-dominated, while the signal PDF is calculated from Monte Carlo simulations. Each individual AGN enters into the signal PDF weighted with the expected relative contribution of the sources, given by their accretion disk flux measured on Earth (see Table~\ref{tab:catalogues}). 

For each direction in the sky, the likelihood function is maximized with respect to the number of signal events $n_s$ and the power-law index of the neutrino spectrum, $\gamma$.
The ratio of the log-likelihood ratio of the best fit hypothesis to the null hypothesis ($n_s = 0$) forms the test statistic. 
The final significance is estimated by applying the same analysis to a large set of scrambled datasets (trials), wherein the right ascensions of the events are randomized but all other event properties are kept fixed. The performance of the method is expressed in terms of discovery potential, defined as the flux required for 50\% of the trials with simulated signal to yield a p-value $\leq 5.73 \times 10^{-7}$ ($5\sigma$).

%%%%%%%%%%%%%%%%%%
%               Performance                %
%%%%%%%%%%%%%%%%%%

\section{Results and perspectives}\label{sec:sens}
\begin{figure}[h]
    \centering
    \includegraphics[width=\textwidth]{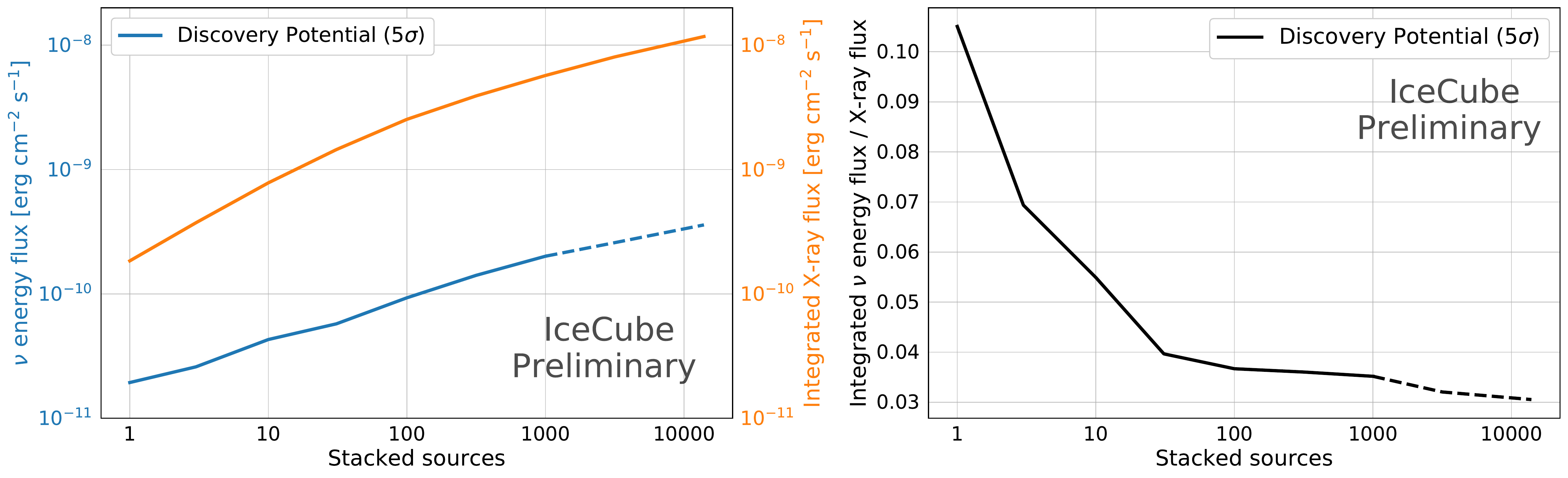}
    \caption{Discovery potential for the radio-selected AGN sample as function of the number of X-ray brightest stacked sources. The left panel shows the integrated energy flux of neutrinos (in blue) and the corresponding integrated X-ray flux (in orange) of the stacked sources. The ratio of the two curves is shown in the right panel, giving the discovery potential per source. The dashed-line indicates that the flux values have been extrapolated for the last two sub-catalogues of stacked sources.}
    \label{fig:dp_sens_subplots} 
\end{figure}

Here, we present preliminary discovery potential estimations for the radio-selected AGN sample. An E$^{-\gamma}$ power law spectrum with $\gamma = 2$ is assumed for each source. In Figure \ref{fig:dp_sens_subplots}, the discovery potential is shown as a function of the number of the stacked sources for each sub-set of the radio-selected AGN sample. Each sub-set contains only the X-ray brightest sources. In the left panel, the integrated neutrino energy flux (blue line) and the integrated X-ray flux (orange line) are shown as a function of the number of stacked sources. The ratio of the two curves is shown in the right panel: it represents the expected neutrino flux per source. Comparing these two plots, it can be noticed that while the discovery potential gets worse as more total flux is required, the same value gets smaller if normalized by the X-ray flux. Therefore, adding sources improves the discovery potential as expected from a stacking analysis. The dashed-line indicates that for the last two sub-samples the values have been extrapolated. In fact, a larger number of trials is required to derive the discovery potential for a number $\mathcal{O}(10^{3})$ of stacked sources. Full results will be shown in a future publication. 

\begin{figure}[h]
    \centering
    \includegraphics[width=\textwidth]{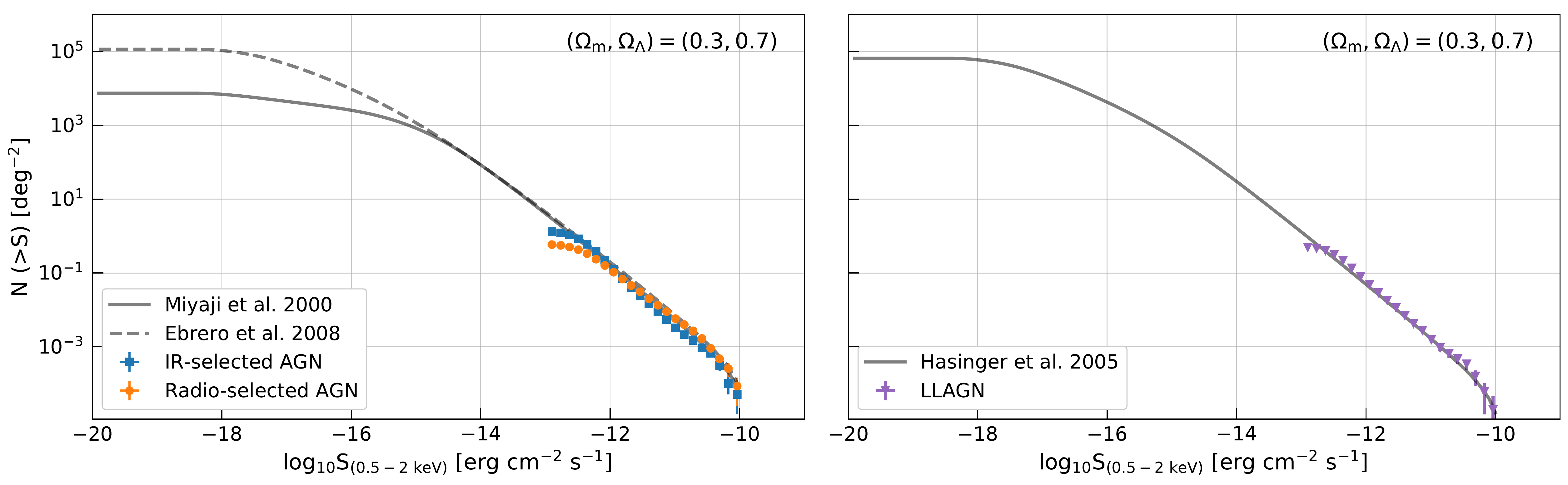}
    \caption{Source count distributions in normalized integral form for sources in the $0.5 - 2$ keV band. Three models (solid lines) derived from X-ray luminosity functions are compared with the sources (points) from the three samples created for this analysis. A cosmological framework with $H_0 = 70 \mathrm{\ km \ s}^{-1}\mathrm{\ Mpc}^{-1}$, $\Omega_{\mathrm{M}} = 0.3$ and $\Omega_{\Lambda} = 0.7$ is assumed.
    }
    \label{fig:lumfunc} 
\end{figure}

\begin{figure}
    % \centering
    \includegraphics[width=0.8\textwidth]{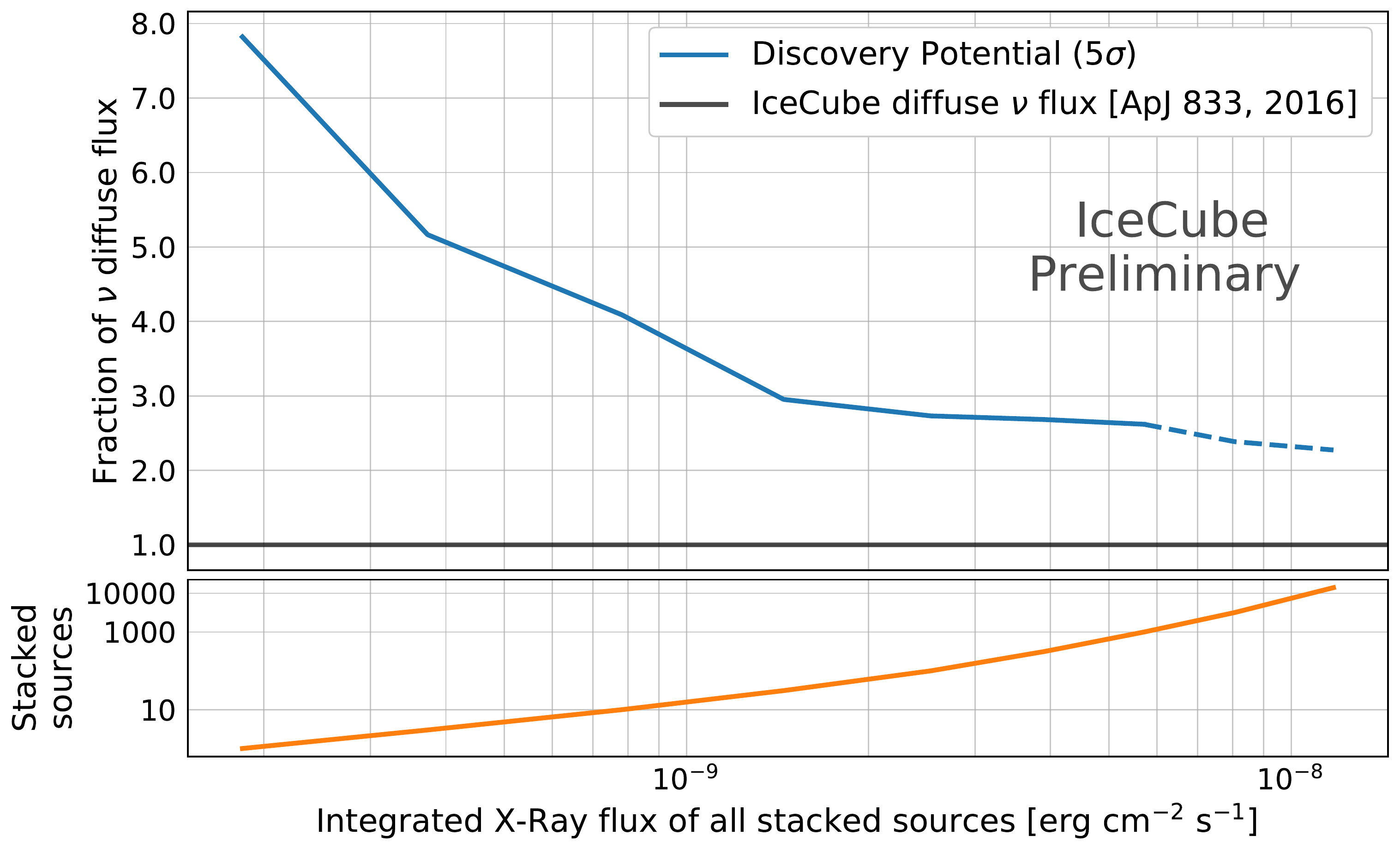}
    \caption{Fraction of diffuse neutrino flux required for a $5\sigma$ discovery at different integrated X-ray flux values (blue line). The corresponding number of stacked sources is plotted below (orange line). The gray line at 1 shows the IceCube diffuse neutrino flux for $\gamma =2$ \cite{2016_diffuse}.} \label{fig:dp_sens}
    \end{figure}

In order to know how many neutrinos are expected from the radio-selected AGN sample, we need to take into account also those sources not included in our selection. For this purpose, the total X-ray flux expected from all AGN has been estimated using the X-ray luminosity function (luminosity-dependent density evolution model; \cite{Miyaji:1999kt, Ebrero:2008hu, Hasinger:2005sb}). Figure \ref{fig:lumfunc} shows the cumulative number of AGN as function of the X-ray flux in the 0.5-2 keV energy range. The expected number of sources derived from the luminosity functions (solid lines) is compared to the number of AGN in the three samples used in this analysis. All samples are scaled by their sky coverage, and the LLAGN sample is also scaled by the efficiency of the selection. 
The total X-ray flux is given by the area below the three theoretical curves: for the radio-selected and IR-selected AGN samples the model from \cite{Miyaji:1999kt} is used (left panel of Figure \ref{fig:lumfunc}), while the LLAGN sample is better described by the model of \cite{Hasinger:2005sb} (right panel of Figure \ref{fig:lumfunc}).
We can then use the total X-ray flux to derive the fraction of the diffuse neutrino flux, defined as the ratio between the integrated neutrino flux normalized over the total flux seen by IceCube, and the X-ray flux of the stacked sources normalized to the total X-ray flux expected at Earth from all the high luminosity AGN. This value is shown in Figure \ref{fig:dp_sens} as a function of the integrated X-ray flux of the stacked sources.  
It tells us what fraction of the diffuse neutrino flux the AGN would have to produce, in order for a discovery to be made. If the fraction is smaller than 1 (i.e. the blue curve in Figure \ref{fig:dp_sens} goes below the gray line), we would be capable of a discovery.  For the normalization of the neutrino flux we use the astrophysical muon neutrino flux from \cite{2016_diffuse} with $\gamma = 2$, which is the same spectral index used for the neutrino spectrum in the analysis.

\begin{figure}
    \floatbox[{\capbeside\thisfloatsetup{capbesideposition={right,center},capbesidewidth=4.5cm}}]{figure}[\FBwidth]
    {\caption{Expected neutrino flux from the radio-selected AGN sample (in light blue) compared to the diffuse neutrino flux with $\gamma = 2$ (solid black line).  Plot adapted from \cite{2016_diffuse}.} \label{fig:diff_flux}}
    {\includegraphics[width=0.65\textwidth]{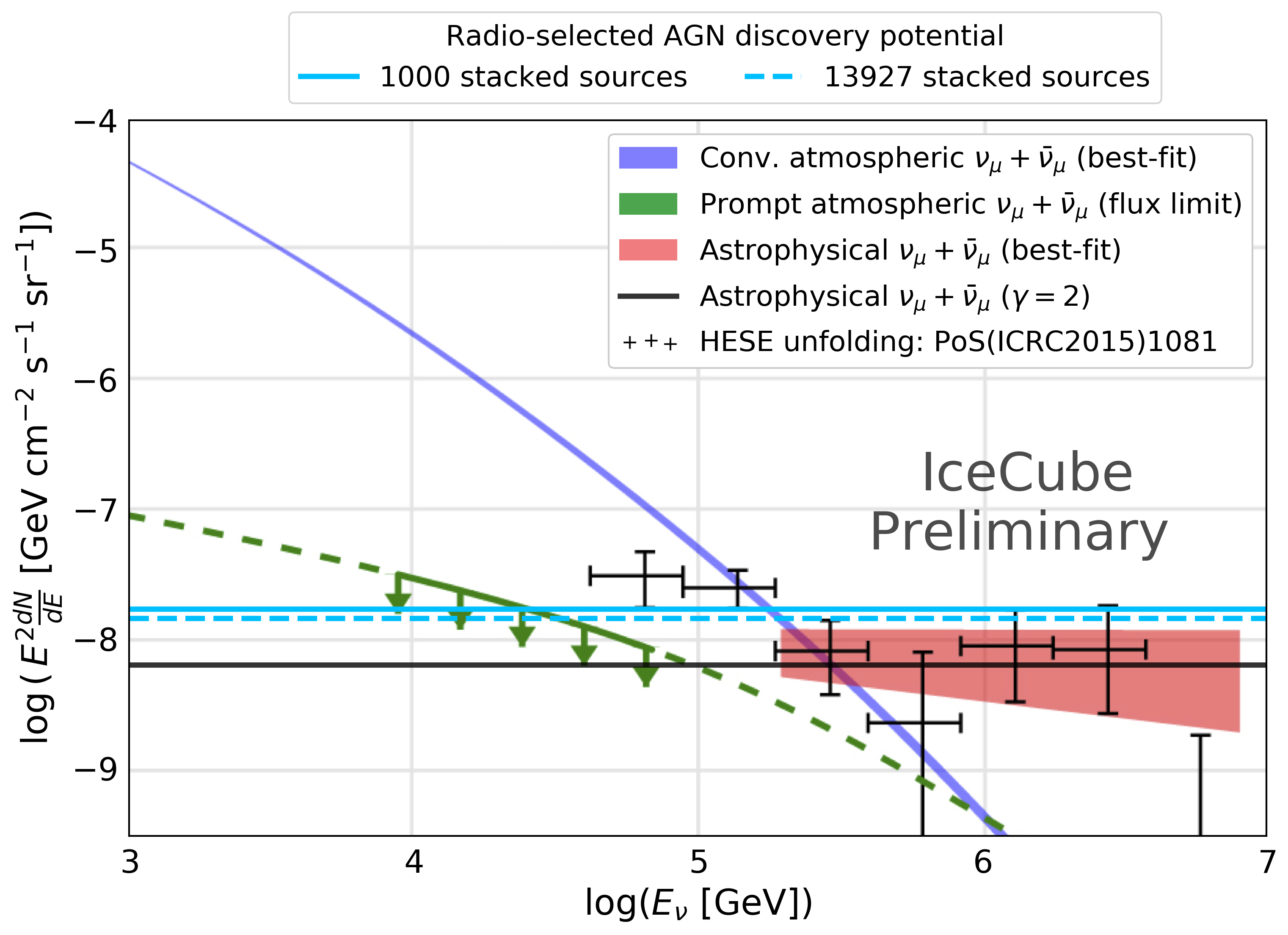}}
\end{figure}

In Figure \ref{fig:diff_flux} the IceCube diffuse flux for $\gamma = 2$ (black horizontal line) is compared to the expected flux from the radio-selected AGN population, for 13,927 and 1,000 stacked sources (light blue lines). They lay above the solid black line, therefore the neutrinos expected from these AGN are not going to saturate the diffuse neutrino flux seen by IceCube. However, we expect the sensitivity of the analysis to be $\sim 3$ times better than the discovery potential, being thus able to go beyond the diffuse limit. We also expect to have lower limits using a softer neutrino spectrum, which is a better description of the IceCube diffuse flux. After running the analysis with the other two samples and with more $\gamma$ values, and unblinding it, the results of the search with corresponding fluxes will be published.

\bibliographystyle{ICRC}
\bibliography{references}
\end{document}